\documentclass[twocolumn]{aastex701}
\usepackage{graphicx}
\usepackage{amsmath}	
\usepackage{amssymb}	
\usepackage{cases}
\usepackage{array,multirow}
\usepackage{upgreek}
\usepackage{textcomp}
\usepackage{lineno}
\usepackage{subcaption}
\usepackage{mhchem}
\usepackage{longtable}
\usepackage{hyperref}
\usepackage[referable]{threeparttablex}
\usepackage{booktabs, dcolumn}
\usepackage{longtable}

\newcommand{\mesa}{\texttt{MESA}}
\newcommand{\tlusty}{\texttt{TLUSTY}}
\newcommand{\synspec}{\texttt{SYNSPEC}}

\shorttitle{Quasi-star Modeling and Spectra}
\shortauthors{Hassan et al.}

\begin{document}

\title{Towards Self-consistent Modeling of Quasi-Stars and their NLTE Spectra: Implications for Little Red Dots}
\correspondingauthor{Jake Hassan}
\email{jakebhassan@gmail.com}

\author[0009-0004-5685-6155]{Jake B. Hassan}
\affil{Department of Physics and Astronomy, Stony Brook University, Stony Brook, NY 11794-3800, USA}
\email{jakebhassan@gmail.com}
\author[0000-0002-3635-5677]{Rosalba Perna}
\affil{Department of Physics and Astronomy, Stony Brook University, Stony Brook, NY 11794-3800, USA}
\email{rosalba.perna@stonybrook.edu}
\author[0000-0003-0936-8488]{Mitchell C. Begelman}
\affil{JILA, University of Colorado and National Institute of Standards and Technology, 440 UCB, Boulder, CO 80309-0440, USA}
\affiliation{Department of Astrophysical and Planetary Sciences, University of Colorado, 391 UCB, Boulder, CO 80309-0391, USA}
\email{mitch@jila.colorado.edu}
\author[0000-0002-2624-3399]{Yan-Fei Jiang}
\affil{Center for Computational Astrophysics, Flatiron Institute, 162 5th Avenue, New York, NY 10010, USA}
\email{jiang@flatitoninstitute.org}
\author[0000-0002-8171-8596]{Matteo Cantiello} 
\affil{Center for Computational Astrophysics, Flatiron Institute, 162 5th Avenue, New York, NY 10010, USA} 
\affil{Department of Astrophysical Sciences, Princeton University, Princeton, NJ 08544, USA}
\email{mcantiello@flatironinstitute.org}
\author[0000-0001-5032-1396]{Philip J. Armitage}
\affil{Department of Physics and Astronomy, Stony Brook University, Stony Brook, NY 11794-3800, USA}
\affil{Center for Computational Astrophysics, Flatiron Institute, 162 5th Avenue, New York, NY 10010, USA}
\email{philip.armitage@stonybrook.edu}

\author[0000-0003-2012-5217]{Taeho Ryu}
\affiliation{JILA, University of Colorado and National Institute of Standards and Technology, 440 UCB, Boulder, CO 80309-0440, USA}
\affiliation{Department of Astrophysical and Planetary Sciences, University of Colorado, 391 UCB, Boulder, CO 80309-0391, USA}
\email{taeho.ryu@colorado.edu}
\affiliation{Max-Planck-Institut für Astrophysik, Karl-Schwarzschild-Straße 1, 85748 Garching bei München, Germany}

\begin{abstract}
Quasi-stars, massive envelopes surrounding rapidly accreting black hole (BH) seeds, are promising candidates for explaining the rapid formation of supermassive black holes in the early Universe and may account for some of the JWST-discovered Little Red Dots (LRDs). We present a self-consistent framework combining stellar evolution calculations with non-local thermodynamic equilibrium (NLTE) spectral synthesis to test this connection. 
Our models drop the polytropic assumptions of previous work: the envelope is described entirely by mixing-length theory, with saturated convection and radiative energy setting the luminosity at an inner boundary a few hundred Schwarzschild radii from the BH. This removes the BH-mass limits of earlier models, and we follow the evolution up to BH-to-total mass ratios approaching unity. 
We then use TLUSTY to compute the first converged NLTE atmospheric models for the surface conditions and primordial hydrogen–helium composition expected in quasi-stars. Our models produce cool photospheres, prominent Balmer breaks, and Balmer and Paschen absorption features. NLTE effects substantially modify line strengths, increasing Balmer absorption depths by $\sim 35-55\%$ relative to LTE models, while having a minor effect on the continuum. Comparison with observed LRD spectral energy distributions shows broad agreement in continuum shape, although our models do not reproduce their commonly observed strong emission lines. Under the quasi-star interpretation, comparisons with individual LRDs suggest characteristic quasi-star masses of approximately $10^5–10^7 M_\odot$. 
These results establish a physically motivated connection between quasi-star evolution and observable spectra, providing a framework for testing quasi-stars as progenitors of LRDs while identifying limitations of current atmospheric models.
\end{abstract}


\section{Introduction} \label{sec:intro}

The existence of supermassive black holes (SMBHs) with masses exceeding $10^9M_\odot$ within the first billion years of the Universe remains an important open problem \citep{Fan2006, Banados2018}. The conventional pathway of Eddington-limited accretion onto stellar-mass black hole (BH) seeds struggles to account for such rapid growth, motivating the exploration of alternative SMBH seed formation channels. Among these, quasi-stars --- massive, radiation-pressure-supported envelopes surrounding accreting BH seeds formed through direct collapse --- have long been recognized as a promising mechanism for rapid early BH growth \citep{Begelman2006, Begelman2008}. In quasi-stars, the central BH accretes at highly super-Eddington rates while the liberated energy is transported outward primarily by convection and radiated near the Eddington luminosity of the entire (BH + envelope) system. Since the envelope is initially much more massive than the BH, this configuration enables the BH to grow rapidly to intermediate masses ($\sim10^3$--$10^4\, M_\odot$ or larger) on Myr timescales, thereby providing plausible progenitors for the SMBHs powering high-redshift quasars.

The observational motivation for the quasi-star paradigm has been revitalized by recent discoveries from the \textit{James Webb Space Telescope} (JWST). Deep extragalactic surveys, including CEERS \citep{Finkelstein2023}, JADES \citep{Rieke2023}, and related programs, have revealed a population of compact, red, high-redshift sources colloquially referred to as ``Little Red Dots'' (LRDs) \citep[e.g.,][]{Labbe2023, Matthee2024, Barro2024, Greene2024, Kokorev2024}. These objects are typically found at redshifts $z\sim 4-10$ and are characterized by compact morphologies, red rest-frame optical colors, and unusual spectral energy distributions (SEDs). In several cases, their SEDs exhibit a pronounced rest-frame optical excess and a steep ultraviolet-to-optical slope, producing a characteristic ``V-shaped'' continuum with a minimum near $\sim3500$~\AA\ in the rest frame \citep{Labbe2023,Matthee2024,Greene2024}.

Spectroscopically, many LRDs display broad Balmer emission lines, including H$\alpha$ and H$\beta$, with full widths at half maximum (FWHM) of order $10^3-5\times10^3$~km~s$^{-1}$ \citep{Maiolino2024,Greene2024}. These line widths suggest the presence of dense, rapidly moving gas analogous to the broad-line regions (BLRs) of active galactic nuclei (AGNs). At the same time, LRD spectra differ significantly from those of classical unobscured quasars: high-ionization ultraviolet lines are often weak or absent, forbidden lines appear suppressed, and some objects exhibit strong Balmer breaks and Balmer absorption features \citep{Maiolino2024,Greene2024}. These properties point toward environments characterized by high gas densities, substantial obscuration, and potentially large optical depths.

A variety of interpretations have been proposed for the LRD population. One class of models identifies LRDs as rapidly accreting, dust-obscured AGNs in which emission from the accretion disk and BLR is partially reprocessed by surrounding gas and dust \citep{Maiolino2024,Greene2024}. Alternative explanations invoke extremely compact and intense starbursts with strong nebular emission contributing substantially to the rest-frame optical flux \citep{Labbe2023}. However, purely stellar models face difficulties in reproducing the observed broad Balmer line widths and equivalent widths without invoking an accreting BH component. Hybrid scenarios involving both star formation and BH accretion have also been considered \citep{Greene2024}.

Quasi-stars have emerged as particularly intriguing candidates for at least a subset of the LRD population \citep{Begelman2025}. Their predicted properties --- large photospheric radii, cool extended envelopes, dense ionized interiors, and high luminosities powered by rapid BH growth --- naturally produce compact, red spectra with strong hydrogen recombination features. Additional support for this interpretation comes from observations of the gravitationally lensed LRD R2211-RX1 \citep{Zhang2025}, which exhibits century-scale photometric variability and a hysteresis loop in the luminosity--temperature plane suggestive of a pulsating stellar-like photosphere. Quasi-star models have been shown to reproduce several of the principal characteristics of this variability \citep{Cantiello2026}.

Significant theoretical progress has recently been made in modeling the internal structure and evolution of quasi-stars. Early calculations by \citet{Ball2011, Ball2012}, based on the Cambridge STARS stellar evolution code, suggested that the embedded BH could not grow beyond a few percent of the total quasi-star mass, casting doubt on the viability of quasi-stars as efficient SMBH seeds. \citet{Coughlin2024} (hereafter CB24) demonstrated that this limitation arose from an unphysical treatment of the inner boundary condition. By replacing the free-fall inner boundary prescription with a ``saturated convection'' model, in which convective transport operates at its maximal efficiency, they showed that the BH can in principle grow to $\sim60\%$ of the total quasi-star mass. Building on this framework, \citet{Hassan2026} implemented the CB24 model within the open-source stellar evolution code \mesa\ \citep{Paxton2011,Paxton2013,Paxton2015,Paxton2018,Paxton2019}, following the time-dependent evolution of quasi-stars under different inner boundary prescriptions and including the effects of eruptive winds. They found that models with dynamic opacity evaluated at the base of the inefficient convection layer terminated at a characteristic BH-to-stellar mass ratio of $M_{\rm BH}/M_\star \simeq 0.33$, largely independent of the initial quasi-star mass.

Despite this progress, several important ingredients remain absent from current quasi-star models. In particular, the \citet{Hassan2026} calculations assumed a fixed adiabatic index $\gamma_o=4/3$ throughout the convective envelope, an approximation that became increasingly inaccurate as gas pressure grew relative to radiation pressure during late evolutionary stages. Furthermore, previous models relied partly on polytropic assumptions and simplified treatments of radiative transport in regions where convection becomes inefficient. At the same time, most spectral synthesis studies of LRDs rely on local thermodynamic equilibrium (LTE) assumptions or simplified nebular prescriptions, despite the fact that the physical conditions inferred for LRDs --- high densities, intense radiation fields, and large scattering optical depths --- naturally place the emitting gas in the regime of non-local thermodynamic equilibrium (NLTE).

Recent theoretical work has begun to explore these effects in more detail. Radiative transfer calculations incorporating resonance, Raman, and Thomson scattering in dense hydrogen gas demonstrate that scattering can significantly broaden Balmer line wings and modify recombination line ratios \citep{Chang2026}. Other studies have shown that collisional excitation in dense gas surrounding high-redshift AGNs can produce intrinsic Balmer absorption features and strong Balmer breaks without requiring stellar populations \citep{Inayoshi2025a}. Semi-empirical models combining standard and unstable accretion disk components have also been proposed to reproduce the V-shaped continua observed in LRDs \citep{Zhang2025disk}.  The recent work of \citet{Martins2026} found that, under their tested range of conditions, supermassive stellar objects can produce LRD-like spectra. However, their approach is agnostic to the nature of the object, particularly the source generating the luminosity. As such, their method alone cannot identify whether an LRD is a quasi-star, a Population III star, or another object.

In this work, we develop a new quasi-star framework that combines improved self-consistent stellar structure calculations with detailed NLTE spectral synthesis. Building on the CB24 and \citet{Hassan2026} models, we remove the reliance on polytropic approximations and instead utilize the full envelope structures computed directly with \mesa, including mixing-length convection and radiative energy transport throughout the interior. We then use the public radiative transfer code \tlusty \ \citep{Hubeny1988,Hubeny1995} to compute NLTE spectra from our quasi-star models.  Our goal is to determine whether quasi-stars can broadly reproduce the main features observed in LRDs, while simultaneously constructing the most physically motivated quasi-star structures to date. This work therefore represents the first quantitative NLTE spectral test of the quasi-star interpretation of LRDs and provides a new framework for connecting JWST observations to models of early SMBH formation.

\section{Modeling of quasi-star evolution}

\subsection{Quasi-star Models in Hassan et al. (2026)}

\citet{Hassan2026} presented a numerical implementation of quasi-star structure and evolution using \mesa, investigating how different inner boundary conditions regulate the growth of the central BH. Their study was based on a modified version of the CB24 model, in which the quasi-star consists of an outer envelope surrounding an inner convective region, with the luminosity approximated by the Eddington limit of the total stellar mass, ($L_{\rm Edd}=4\pi G M_\star c/\kappa$). Unlike CB24, which assumed a constant electron-scattering opacity, ($\kappa=\kappa_{\rm es}=0.34~{\rm cm}^2~{\rm g}^{-1}$), \citet{Hassan2026} computed the opacity dynamically at the base of the inefficient convection (IC)  layer, where energy transport via radiative diffusion becomes comparable to  convective transport.  
They further modeled the envelope directly with \mesa, rather than evolving only the core and attaching a polytropic envelope.

Despite these improvements, several aspects of the implementation remained tied to assumptions inherited from CB24. In particular, the governing equations were derived under the assumption that the envelope closely resembles a $\gamma=4/3$ adiabatic polytrope, an approximation that breaks down once the BH mass exceeds $M_{\rm BH}\sim 0.3~M_\star$. Furthermore, \citet{Hassan2026} adopted the CB24 prescription for the inner boundary radius, $R_i$, below which convection is assumed to be saturated. In CB24, this radius was identified as the location where the entropy gradient becomes positive ($ds/dr>0$) in the analytic envelope solution. However, this criterion no longer has a clear physical motivation when the envelope is modeled with \mesa\ and standard Mixing Length Theory (MLT), since the entropy gradient remains negative throughout the envelope. Consequently, the choice of $R_i$ is no longer uniquely determined and must instead be treated as a free parameter.

With their modeling, \citet{Hassan2026} found that the BH grows to approximately $M_\mathrm{BH}/M_\star \simeq 0.33$ at termination, largely independent of the initial quasi-star mass. This value is roughly a factor of two lower than the analytical upper limit of $M_\mathrm{BH}/M_i \simeq 0.62$ predicted by CB24, a discrepancy they traced to the thin-atmosphere approximation adopted in that work, whereby the mass at the convective boundary $M_{\rm crit}$ was assumed to equal $M_\star$ at all times (similarly to
the quasi-star implementation of \citealt{Santarelli2025}). In the more realistic treatment of \citet{Hassan2026}, $M_{\rm crit}$ was found to lie well below $M_\star$ throughout the evolution, reducing the maximum possible size of the black hole. The photospheric temperatures of their models decreased monotonically with time, settling in the range $\sim$4500--4800\,K for most of the quasi-star's lifetime, with more massive models exhibiting slightly higher effective temperatures. \citet{Hassan2026} also explored the effect of eruptive winds, implemented following the prescription of \citet{Cheng2024}. They found that, despite severe mass depletion, the final $M_\mathrm{BH}/M_\star$ ratio remained close to the wind-free value, reflecting the self-similar nature of the CB24 model.

\subsection{Updates to the \texttt{MESA} quasi-star model}

Our latest models run on version r26.4.1 of \mesa, and the inlists and code for them are available in a public repository.\footnote{\url{https://github.com/JakeBHassan/Quasistar-Models}} In our current approach, we seek to use fewer analytical assumptions, keeping the model simple and more consistent with \mesa's mixing length theory. We use the following inner boundary conditions:
\begin{equation}
    R_i = N r_s(M_{\rm BH})
\end{equation}
\begin{equation}
    M_i \approx M_{\rm BH}
\end{equation}
\begin{equation}
    L_i = L_{\rm sat}(R_i) + L_{\rm rad}(R_i).
\end{equation}
Our $R_i$ is chosen to be a multiple of the BH's Schwarzschild radius $r_s(M_{\rm BH})=2GM_{\rm BH}/c^2$, with the scale factor $N$ set to the lowest value we can consistently get converged \mesa \ models from. In this setup, we found that $N=300$ works across a wide range of quasi-star masses. For comparison, the Bondi radius of the BH lies at $N\approx 830$, while the inner radius used in \cite{Hassan2026} is on the order of $N \sim 2\times 10^5$; we are thus able to model significantly deeper into the quasi-star. The Tolman--Oppenheimer--Volkoff (TOV) general relativistic correction is included for good measure, although it has a very minor impact even for $N \sim 10$.

For simplicity, $M_i$ is assumed to equal $M_{\rm BH}$ in our model. This approximation is increasingly valid as $N$ is lowered, and differences become insignificant for $N \lesssim 300$. In principle, $M_i$ could be exchanged for an analytic estimate of the matter contained within $R_i$, though we did not deem this necessary for our purposes for the reason above.

Like CB24, we take the convective luminosity at $R_i$ to be the saturated limit, proportional to the enthalpy advected at the medium's sound speed:
\begin{equation}
F_{\rm sat} \propto \frac{\gamma}{\gamma-1} pc_s\,,
\end{equation}
where $\gamma$ is the adiabatic index, $p$ is the pressure and $c_s$ is the speed of sound. The convective flux at the interior boundary is then assumed to be $L_{\rm conv}(R_i)=L_{\rm sat}(R_i)=4\pi R_i^2 F_{\rm sat}(R_i)$. We define a convective efficiency parameter $\beta\in(0, 1]$ for which
\begin{equation}
F_{\rm sat} = \beta \frac{\gamma_{\rm rel}}{4\gamma_{\rm rel}-3} pc_s
\end{equation}
where $\gamma_{\rm rel}=\gamma/(4/3)$. This definition is similar to that used in previous papers, while also removing the assumption $\gamma=4/3$ from this model (as we instead obtain $\gamma$, $p$ and $c_s$ from the base of \mesa's envelope). $\beta$, as before, is a parameter that must be provided by the user; we assume it to be 0.1 for our own simulations.

Our luminosity also contains a notable correction compared to previous models. CB24 (and thus also \citealt{Hassan2026}) assumed the radiative contribution to luminosity to be insignificant compared to $L_{\rm conv}$, and therefore approximated $L=L_{\rm conv} = L_{\rm sat}$ to simplify the analytic derivations. In our new model, we instead impose the boundary condition $L_i=L_{\rm sat} (R_i) + L_{\rm rad}(R_i)$, where $L_{\rm rad}(R_i)$ is obtained from \mesa \ at the innermost zone. While the condition $L_{\rm conv} = L_{\rm sat}$ is still imposed, we now incorporate a nonzero $L_{\rm rad}$ into the total luminosity.
The radiative luminosity was found to be within 0.1\% of the local Eddington limit $L_{\rm Edd}(R_i)$ at all times, as expected of a region dominated by radiation pressure. Because $L_{\rm Edd}(R_i) \propto M_i$, radiative transport becomes increasingly relevant at later times, and by the end of life it comprises nearly half of the luminosity. We also note that we no longer need to make the approximation $L=L_{\rm Edd}(R_\star)$ anywhere in our equations, unlike \cite{Hassan2026}; this removes any need to decide on the opacity to use for calculating the luminosity (i.e. $\kappa_{\rm es}$ vs. $\kappa_{\rm IC}$).

As in CB24 and \cite{Hassan2026}, the BH accretes matter at a rate proportional to the current inner boundary luminosity:
\begin{equation}
L_i=\epsilon \dot{M}_{\rm in} c^2=\epsilon' \dot{M}_\mathrm{BH} c^2\,,
\end{equation}
where $\epsilon$ is the radiative efficiency parameter, $\epsilon' = \epsilon/(1-\epsilon)$, and $\dot{M}_{\rm in}$ is the mass inflow rate across the base of the envelope. Here we use the conventional value of $\epsilon \sim 0.1$, but keeping in mind that this is a model parameter which may vary. Its value  determines how efficiently the BH accretes matter. Because the state of the quasi-star is entirely determined by the current value of $M_{\rm BH}$, $\epsilon$ reflects an uncertainty in our knowledge of the quasi-star's lifespan; in other words, it determines the speed of the evolution, but does not affect the quasi-star's structure throughout the evolution.

The choice to model the envelope self-consistently with \mesa \ comes with multiple benefits. It avoids having to make any assumptions or idealizations about the envelope structure; rather than assuming it resembles a polytrope with $\gamma=4/3$, the envelope is now described entirely with mixing length theory, such that its innermost zone agrees with the equations for the interior. Additionally, as stated before, our usage of MLT means that tying the location of the inner boundary to an entropy gradient would be meaningless. Aside from allowing us to model much deeper and closer to the BH, the fact that our inner boundary is no longer tied to the entropy gradient means we no longer run into the upper limit of $M_{\rm BH}/M_\star \sim 0.62$ and $M_{\rm BH}/M_\star \sim 0.33$ found in CB24 and \cite{Hassan2026}, respectively. Indeed, because we do not face this analytical constraint, our model covers the entire lifetime of the quasi-star up to $M_{\rm BH}/M_\star \sim 0.99$ (the only exception being our $10^7 M_\odot$ models, which cover up to $M_{\rm BH}/M_\star \sim 0.95$).

1D MLT produces a density inversion near the surface of our models (Section~\ref{sec:results}). To test its effect, we also compute one model with \mesa's superadiabatic reduction (SR; \citealt{Jermyn2023}).
Like its predecessor MLT++ \citep{Paxton2013}, SR is a numerical prescription that stands
in for multi-dimensional effects (e.g.\ porosity) that let near-Eddington envelopes carry
more flux than 1D MLT allows. Where the local radiative Eddington factor
$\Gamma = 4(1-\beta_{\rm g})\nabla_T$ exceeds a threshold $\Gamma_{\rm lim}$ or the
density-inversion threshold $\Gamma_{\rm inv} = 4(1-\beta_{\rm g})/(4-3\beta_{\rm g})$,
with $\beta_{\rm g} = P_{\rm gas}/P$, SR replaces $\nabla_{\rm rad}$ by
$\nabla_{\rm L} + (\nabla_{\rm rad}-\nabla_{\rm L})/f_{\rm SR}$ and re-solves MLT. Lowering $\nabla_T$ lowers $L_{\rm rad}$, so more of the flux is assigned to convection.
The factor scales as $f_{\rm SR} - 1 \propto \alpha_{\rm SR}/\sqrt{\beta_{\rm g}}$, so it
becomes large in radiation-dominated layers, and its strength $\alpha_{\rm SR}$ is a free
parameter. We use $\Gamma_{\rm lim} = 0.5$ (\texttt{superad\_reduction\_Gamma\_limit};
the default) and $\alpha_{\rm SR} = 25$ (\texttt{superad\_reduction\_Gamma\_limit\_scale};
default 5), with all other SR controls at their defaults. Our inner boundary condition does not account for SR, so models with SR enabled are not fully self-consistent with our assumptions.

Lastly, compared to the code in \citet{Hassan2026}, we have also found optimizations to make the model run several dozens of times faster. This allows it to quickly run to completion on personal computers, as opposed to requiring supercomputers. Our code additionally includes bug fixes to improve numerical stability, allowing for the simulation of high-mass quasi-stars ($\ge10^6 M_\odot$).

\section{Spectral modeling}

We use version 208 of the atmospheric modeling code \tlusty \ \citep{Hubeny1988,Hubeny1995, Hubeny2021} to generate NLTE atmospheric models and synthetic spectra for our quasi-stars. \tlusty\ is a 1D, static, plane-parallel modeling code. In contrast, \citet{Martins2026} used CMFGEN to compare LRDs with supermassive star atmospheres; their models are spherical and dynamic, and use a nonzero wind mass loss. Although we cannot account for those aspects, our use of \tlusty\ pairs well with \mesa; while \mesa\ can describe stellar expansion and winds, it is still a hydrostatic code, and there are limitations to how reliable it can be for detailing the surface conditions of a dynamic envelope. For our purposes, assuming constant mass in \mesa\ and constructing a \tlusty\ model based on its surface is a reliable way of deriving a quasi-star spectrum from first principles, and can be viewed as a representation for a quasi-star with low mass loss.

The effective temperature, gravitational acceleration and composition are used as input parameters for our atmospheres. \tlusty \ accepts an initial model as a starting point and iterates upon it to achieve radiative equilibrium, hydrostatic equilibrium and charge conservation. Achieving convergence at low surface gravities ($\log g<0$) is particularly difficult; for example, \cite{Liu2026} obtained LTE models in this regime by modifying the source code to introduce additional assumptions. We achieved this in NLTE without having to modify \tlusty \ by identifying a sequence of steps that works for several tested conditions. This broadly follows the recommended chain of phases from the \tlusty \ documentation: gray atmosphere $\rightarrow$ LTE with lines $\rightarrow$ NLTE continuum-only $\rightarrow$ NLTE with lines. Each of these phases is individually split into steps that gradually change the parameters to fit our desired conditions (e.g.~mixing length parameter), so that the model remains numerically stable. We leave the microturbulent velocity at zero, so the only sources of line broadening are Stark and thermal Doppler broadening.

Because quasi-stars are expected to have near-zero metallicities, our main \mesa \ and \tlusty \ models only account for hydrogen and helium, at proportions consistent with Big Bang Nucleosynthesis ($Y=0.2467$; \citealt{Skillman2026}). We therefore provide the first zero-metallicity models at the extreme surface conditions of quasi-stars, across both LTE and NLTE. As a point of comparison, we also tested models with $[\rm{M/H}]=-2$ and $[\rm{M/H}]=-1$, which are the compositions used in prior works \citep{Liu2026, Martins2026}. We use version 54 of \synspec \ to generate spectra for these models, as \tlusty \ alone approximates metal levels to achieve convergence. As these are higher than the predicted metallicities of quasi-stars and are thus of lower priority, we conceded certain physical details for the sake of computational speed. Firstly, we created them using partial opacity tables, so only seven species are treated in NLTE (H, He, Na, Mg, Si, Ca, Fe) while the rest use LTE opacities. Secondly, \tlusty \ currently does not support using the complete linearization (CL) scheme for NLTE models with opacity tables \citep{Hubeny2021}. These models must therefore use the extended Rybicki scheme, in which the radiation field and temperature structure are linearized, while the NLTE level populations, charge conservation and the hydrostatic equilibrium are updated between linearization steps through ALI-accelerated iterations. Our metal-free models do not have this limitation and were obtained using CL. Thirdly, for easier convergence, we did not account for charge exchange.


\section{Results} \label{sec:results}

\subsection{Quasi-star Evolutionary Model}
Upon initialization of our model as a protostar, we introduce a central BH, initially comprising $10^{-3}$ of the total system mass, by modifying the inner boundary conditions. For the purposes of this work, we assume constant mass when finding spectral models at hydrostatic equilibrium; a detailed analysis of the influence of wind mass loss on the evolution and structure of quasi-stars can be found in \citet{Hassan2026}.

\begin{figure*}[ht!]
\centering
\includegraphics[width=\linewidth]{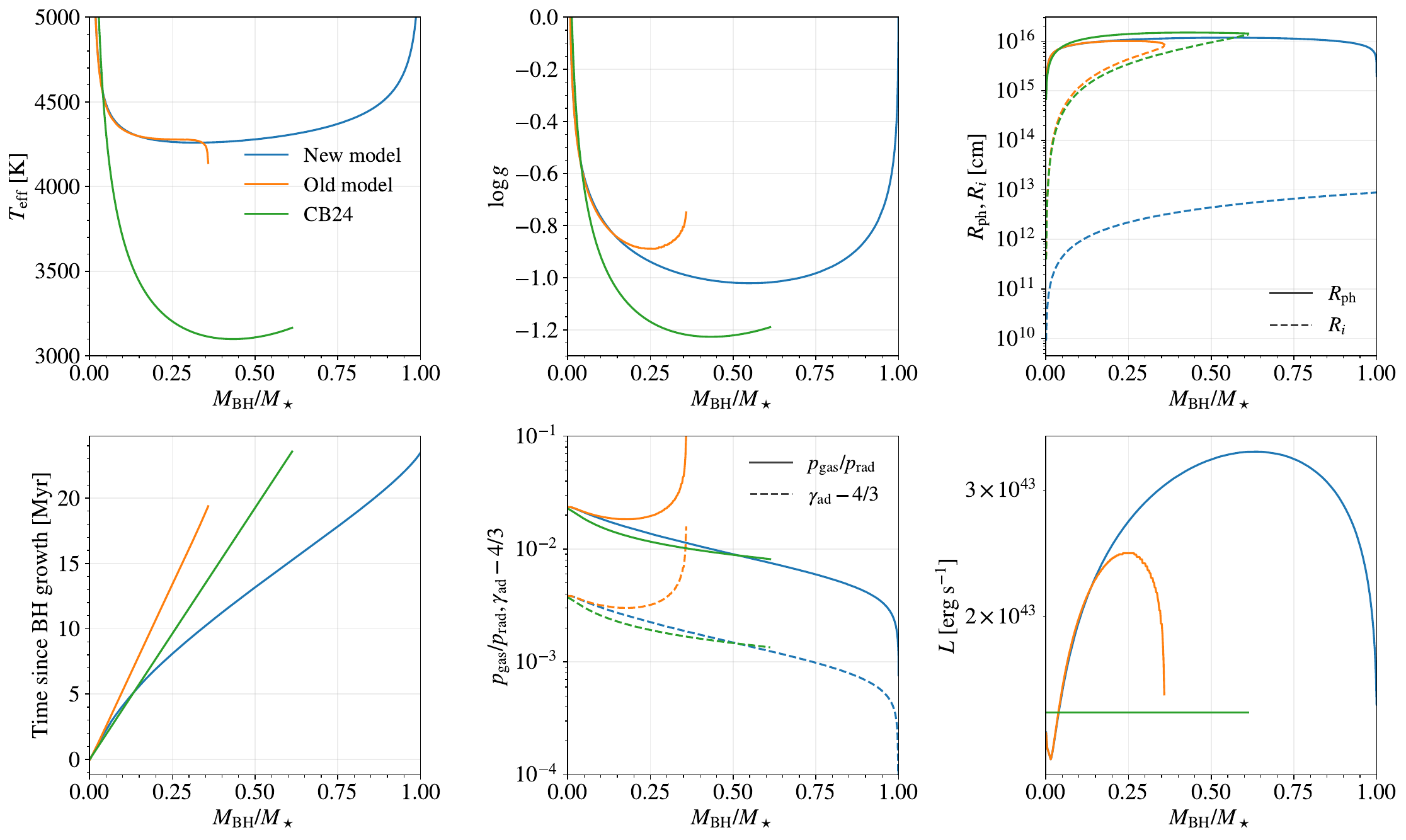}
\caption{Comparison of several properties between the new \mesa \ model, the older model from \citet{Hassan2026} and the analytic predictions of CB24. Models are taken to have the same parameters, with $M_\star = 10^{5} M_\odot$. The $p_{\rm gas}/p_{\rm rad}$ and $\gamma_{\rm ad}$ values are obtained at the middle of the envelope. Note that the models have different definitions of the inner radius BC $R_i$.}
\label{fig:model_comparison}
\end{figure*}

Figure \ref{fig:model_comparison} shows the evolution of various physical properties for a $10^5 M_\odot$ quasi-star, comparing the results of our new model with the polytrope-based ones from \cite{Hassan2026} and the purely analytic model from CB24. The surface properties of our current model largely agree with those of \cite{Hassan2026}, although deviations begin to manifest when the latter nears its limiting black hole mass. Interestingly, despite our new model making no polytropic assumptions in its derivation, its mid-envelope values of $p_{\rm gas}/p_{\rm rad}$ and $\gamma_{\rm ad}$ are closer to the predictions of CB24's polytropic model than \cite{Hassan2026}; namely, $\gamma_{\rm ad}$ tends toward 4/3 over time in our new model, whereas the old model asymptotically diverged from it.

\begin{figure*}[ht!]
\centering
\includegraphics[width=\linewidth]{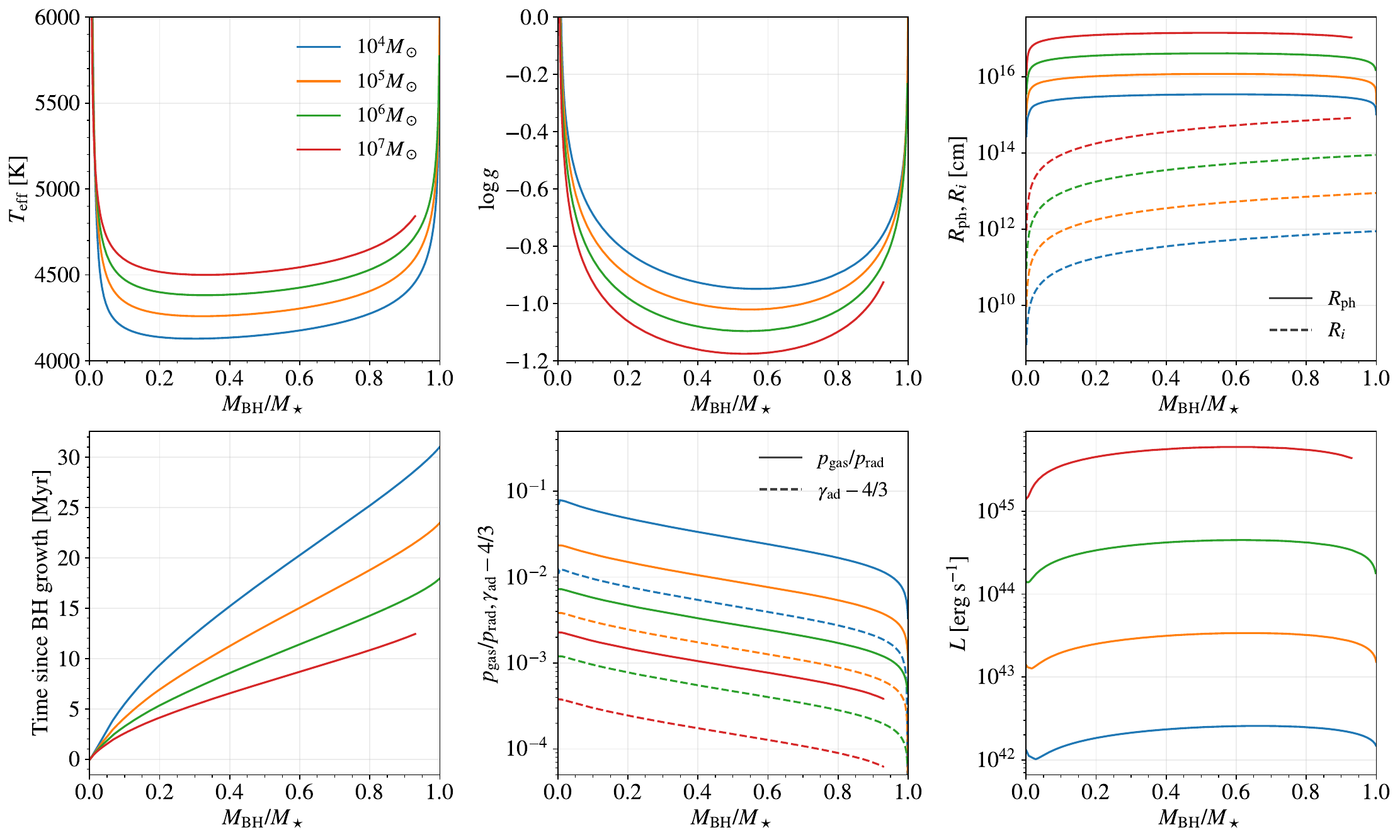}
\caption{Comparison of the properties of our new quasi-star models across a range of masses. All models in this figure use $\alpha_{\rm MLT}=1$ and \texttt{atm\_T\_tau\_opacity = fixed}.}
\label{fig:masses}
\end{figure*}

Figure \ref{fig:masses} shows a similar plot, comparing the evolution for different quasi-star masses between $10^4-10^7 M_\odot$. An order of magnitude increase in quasi-star mass generally leads to an increase in effective temperature by $\Delta T_{\rm eff} \sim 100-150$. All masses demonstrate a similar evolution as functions of $M_{\rm BH}/M_\star$: the effective temperature and surface gravity drop at the start of the model, remain fairly constant for most of the quasi-star's lifetime, and then increase near the end. There is thus potential to identify the mass of the inner black hole from the surface properties of quasi-stars, although our current models suggest it is a relatively weak correlation and would not be one-to-one with mass. Additionally, the physical timescale of the evolution depends on the total mass, with more massive quasi-stars evolving faster and thus having shorter lifespans.

The surface properties of our quasi-stars have little dependence on the convective efficiency $\beta$, and as stated earlier, no dependence on the radiative efficiency $\epsilon$. Indeed, aside from the BH's mass, the properties of the saturated interior appear to have little influence on the envelope's outer regions; this could explain why the different interior treatments of our new and old model lead to similar surface properties, except for where the older model approaches its BH mass limit. In contrast, the surface is more directly influenced by the settings used for the envelope's gas, in particular the mixing length parameter $\alpha_{\rm MLT}$, the treatment of opacity, and the choice of whether to enable SR. $\alpha_{\rm MLT}$ is the ratio of the mixing length to the local pressure scale height. It is usually calibrated against the Sun or against 3D simulations of surface convection in low-mass stars, giving $\alpha_{\rm MLT}\approx 1.8$--$2$. Neither calibration applies to radiation-dominated quasi-star envelopes, so we treat it as a free parameter and bracket it with $\alpha_{\rm MLT}=1$ and $2$. A higher value of $\alpha_{\rm MLT}$ increases the efficiency of convection, leading to higher $T_{\rm eff}$ and lower $\log g$ in our models. For opacity, \mesa \ provides the option of either fixing the opacity everywhere to its value at the surface (\texttt{atm\_T\_tau\_opacity = fixed}) or using a varying opacity consistent with the structure of the atmosphere (\texttt{atm\_T\_tau\_opacity = varying}). \citet{Hassan2026} was only able to obtain numerical stability with the former, whereas we are now able to use either in our current model. In the absence of SR, our models with ``varying'' generally have effective temperatures $\sim 300-400 \ \rm K$ higher compared to using ``fixed''. We ultimately produced 16 non-SR \mesa \ models for analysis, across a grid of four masses ($10^4 -10^7 M_\odot$), two $\alpha_{\rm MLT}$ values (1.0 and 2.0), and two opacity treatments (``fixed" and ``varying"). For our discussions, we will refer to the ``fiducial model'' as the non-SR \mesa \ model with $M_\star = 10^6 M_\odot$, ``fixed'' opacity, and $\alpha_{\rm MLT}=1$.

\begin{figure*}[ht!]
\centering
\includegraphics[width=\linewidth]{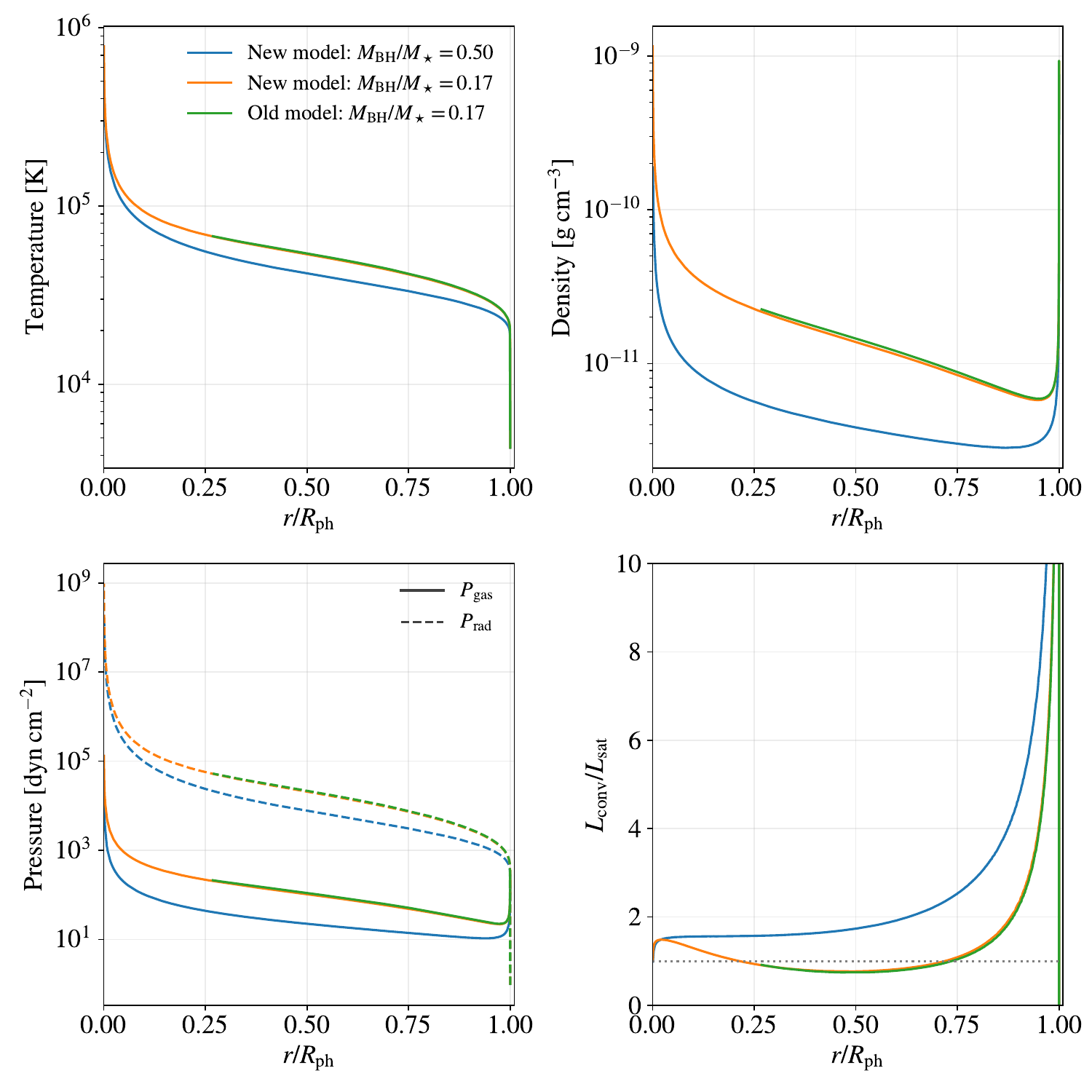}
\caption{Several structural profiles for our fiducial quasi-star. We show the results at $M_{\rm BH} = 0.5 M_\star$, as well as the comparison between our new and old models at $M_{\rm BH} \approx 0.17 M_\star$.}
\label{fig:MESAprofile}
\end{figure*}

In Figure \ref{fig:MESAprofile}, we show the temperature, density, pressure and convective luminosity of the fiducial model at $M_{\rm BH} = 0.5 M_\star$. We also compare its structure to the older \citet{Hassan2026} approach, at half the termination age of the older model ($M_{\rm BH} \approx 0.17 M_\star$). We see close agreement at these earlier times, with a density inversion forming near the surface. The temperature and density of the quasi-star model also declines between these two times.

CB24 predicts saturated convection deep into the interior and sub-saturated convection in the envelope; we see that $L_{\rm conv}$ exceeds $L_{\rm sat}$ throughout the envelopes in Figure \ref{fig:MESAprofile}. \mesa \ does not provide a means of applying a limit on the convective luminosity, likely because of how rarely it would be needed for normal stars. In both \cite{Hassan2026} and our newer model, we impose $L_{\rm conv} = L_{\rm sat}$ at the innermost zone, but we cannot impose $L_{\rm conv} \leq L_{\rm sat}$ throughout the entire envelope. At earlier times, most of the quasi-star's convective luminosity remains below the cap, with the exception of the outermost layers nearing the density inversion. Above a certain threshold ($M_{\rm BH}/M_\star \sim 0.24$), our current model exceeds the cap everywhere except the innermost zone (due to us imposing it). The older model, with its inner radius set by the entropy gradient of the polytropic model, imposes the $L_{\rm conv} = L_{\rm sat}$ constraint at a much larger radius than our newer one; once the newer one exceeds $L_{\rm sat}$ everywhere, differences between the models start to become noticeable. By $M_{\rm BH}/M_\star \sim 0.33$, the older model attempts to impose this constraint near the surface, where the density inversion (and therefore the spike in $L_{\rm conv}/L_{\rm sat}$) occurs. It is likely unable to find a physically viable solution beyond this point, leading to its divergences and early termination.

At $\beta=0.1$, $L_{\rm conv}/L_{\rm sat}\leq2$ for $r/R_\star < 0.4$ at all times, increasing beyond that at shallower depths. This deviation can be reduced by using lower values of $\beta$, but to do so would not be a physically satisfactory solution: the theory is that the envelope should adjust so that it does not exceed $L_{\rm sat}$ (regardless of the value of $\beta$, which is a free parameter), so it is meaningless to constrain $\beta$ to match a non-flux limited envelope. Although we tested extending the code to introduce a flux limiter ourselves, this frequently led to unphysical or highly unstable envelope structures. For the purposes of getting a physically motivated $L_i$, we settle for imposing the constraint on the innermost zone, leaving the investigation of an envelope flux limiter to future works.

\begin{figure*}[ht!]
\centering
\includegraphics[width=\linewidth]{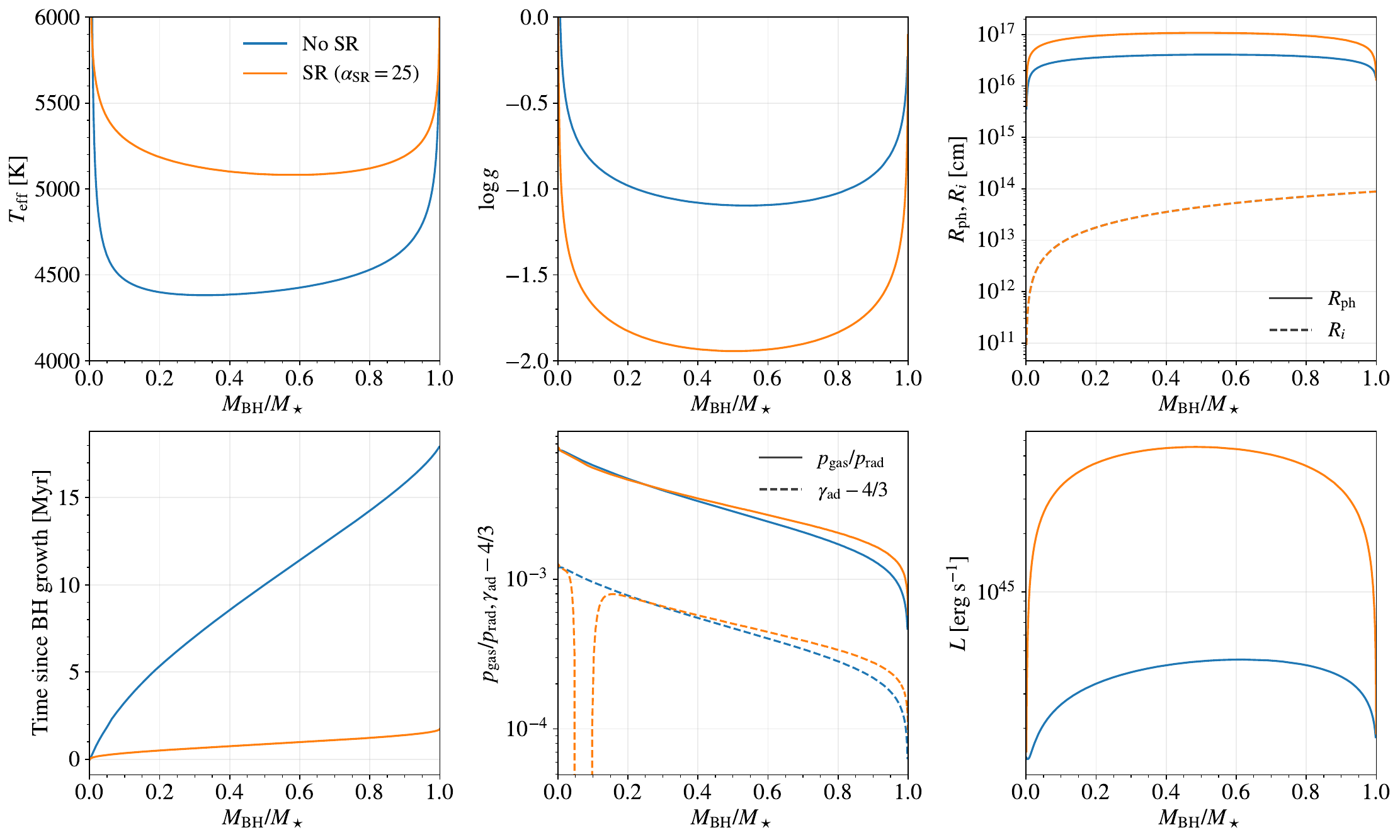}
\caption{Comparison of properties for the fiducial model, both with and without superadiabatic reduction enabled. Both models use $M_\star = 10^6 M_\odot$, $\alpha_{\rm MLT}=1$, and a ``fixed" opacity.}
\label{fig:SRevolution}
\end{figure*}

To test the influence of SR, we computed a model matching our fiducial model but with SR
enabled ($\alpha_{\rm SR} = 25$), chosen large enough to clearly suppress the surface
density inversion. SR models require the ``fixed'' atmosphere: with ``varying'', which
has no radiative-force term, the SR model drifts to an unphysical super-Eddington photosphere.

Figure \ref{fig:SRevolution} compares the non-SR fiducial model with its SR counterpart. SR substantially increases the luminosity (primarily through the convective component) and radius of the quasi-star. The surface $\log g$ is much lower, while the effective temperature increases by several hundred kelvins.

\begin{figure*}[ht!]
\centering
\includegraphics[width=\linewidth]{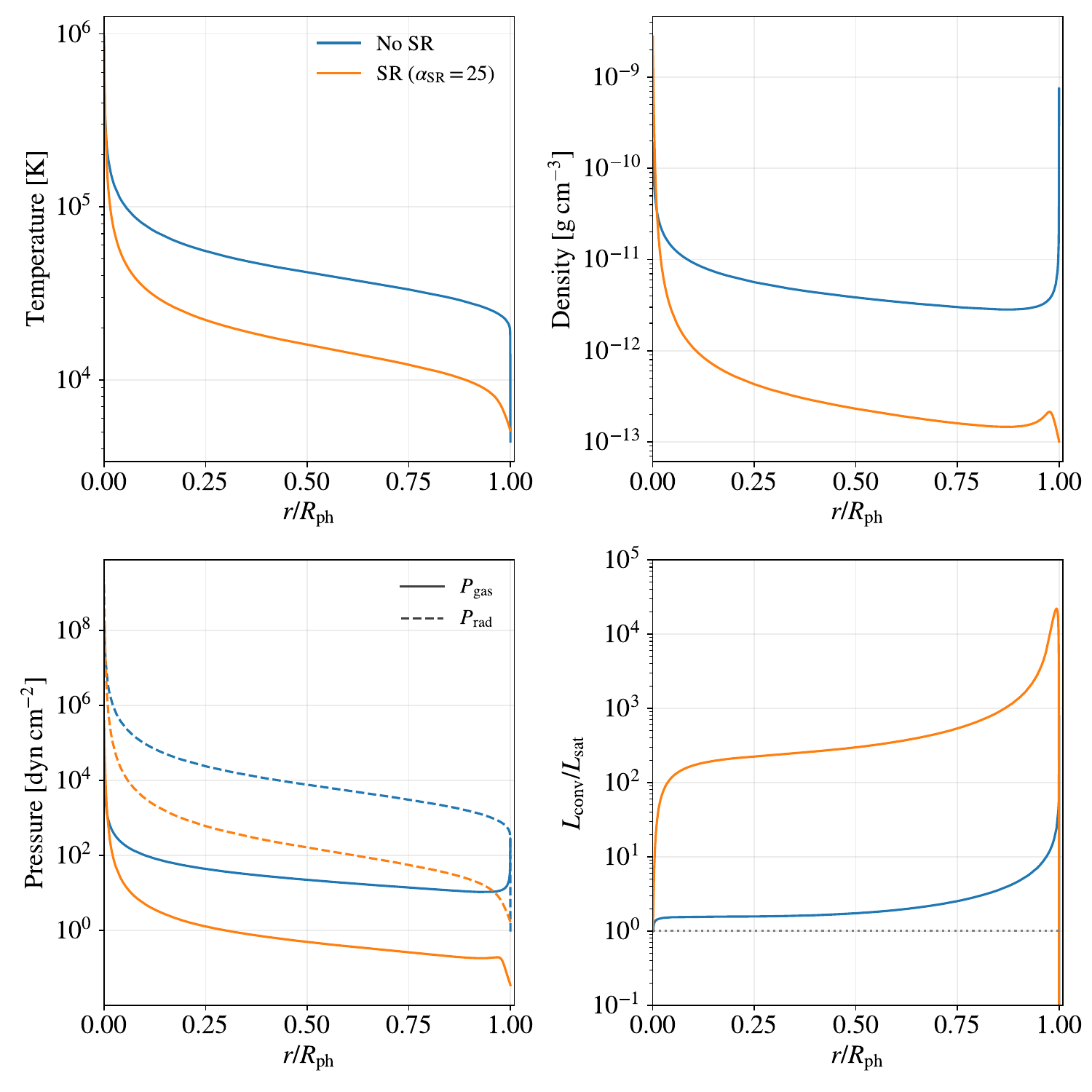}
\caption{Similar plot to Figure \ref{fig:MESAprofile}, but instead comparing the structures at $M_{\rm BH} = 0.5 M_\star$ for our fiducial model with and without SR enabled.}
\label{fig:SRprofile}
\end{figure*}

In Figure \ref{fig:SRprofile}, we compare the midlife ($M_{\rm BH}/M_\star = 0.5$) profiles for both of these models. SR lowers the density and pressure profiles by roughly an order of magnitude, and the inversion at the surface is now smoothed into a narrow bump. The temperature profile of the SR model is likewise lower than the non-SR model throughout the entire star, with the exception of the surface: the non-SR model's temperature sharply declines at the density inversion, causing its effective temperature to be lower than that of SR. Additionally, our non-SR model has $L_{\rm conv}/L_{\rm sat}<2$ in its lower half, with this ratio rising toward the surface. In contrast, the SR model has $L_{\rm conv}/L_{\rm sat} > 100$ throughout almost all of its envelope. This is a direct consequence of SR, which lowers $\nabla_T$ and assigns to convection the flux radiation no longer carried by radiation. Since $L_{\rm sat}$ is the enthalpy flux advected at the sound speed, the SR envelope carries far more convective flux than subsonic convection can. We therefore treat the SR model as a test of the surface density inversion, not as a
physical quasi-star structure.

\subsection{Quasi-star Spectra}

\begin{table}
\begin{center}
    \begin{tabular}{c c c | c c c c} 
 \hline
    $M_\star$ & Opacity & $\alpha_{\rm MLT}$ & $T_{\rm eff}$ & $\log g$ & Balmer & $\Delta R / R_{\rm ph}$ \\ $(M_\odot)$ & type & & $(\rm K)$ & &
  break & $(\times 10^{-3})$ \\ \hline
    \hline
    $10^{4}$ & fixed & 1 & 4151 & $-0.95$ & 2.89 & 2.3 \\
    $10^{5}$ & fixed & 1 (2.09) & 4278 & $-1.02$ & 2.76 & 1.0 \\
    $10^{6}$ & fixed & 1 & 4400 & $-1.10$ & 2.67 & 0.4 \\
    $10^{7}$ & fixed & 1 & 4518 & $-1.18$ & 2.58 & 0.2 \\
    \hline
    $10^{4}$ & fixed & 2 & 4394 & $-1.13$ & 2.67 & 3.6 \\
    $10^{5}$ & fixed & 2 & 4526 & $-1.20$ & 2.58 & 1.5 \\
    $10^{6}$ & fixed & 2 & 4654 & $-1.28$ & 2.52 & 0.6 \\
    $10^{7}$ & fixed & 2 & 4778 & $-1.36$ & 2.48 & 0.2 \\
    \hline \hline
    $10^{4}$ & varying & 1 & 4444 & $-0.89$ & 2.58 & 2.3 \\
    $10^{5}$ & varying & 1 & 4580 & $-0.96$ & 2.49 & 1.2 \\
    $10^{6}$ & varying & 1 & 4714 & $-1.04$ & 2.44 & 0.5 \\
    $10^{7}$ & varying & 1 & 4843 & $-1.12$ & 2.41 & 0.2 \\
    \hline
    $10^{4}$ & varying & 2 & 4714 & $-1.07$ & 2.44 & 5.1 \\
    $10^{5}$ & varying & 2 & 4860 & $-1.15$ & 2.40 & 1.9 \\
    $10^{6}$ & varying & 2 & 5014 & $-1.22$ & 2.38 & 0.7 \\
    $10^{7}$ & varying & 2 (18.5) & 5191 & $-1.30$ & 2.23 & 0.3 \\
    \hline
\end{tabular}
\caption{\mesa \ and \tlusty \ properties for our metal-free non-SR models at $M_{\rm{BH}}/M_\star=0.5$. Values of \texttt{HMIX0} that differ from the \mesa \ $\alpha_{\rm MLT}$ are included in parentheses.}
\label{tab:grid}
\end{center}
\end{table}

\begin{figure*}[ht!]
\centering
\includegraphics[width=\linewidth]{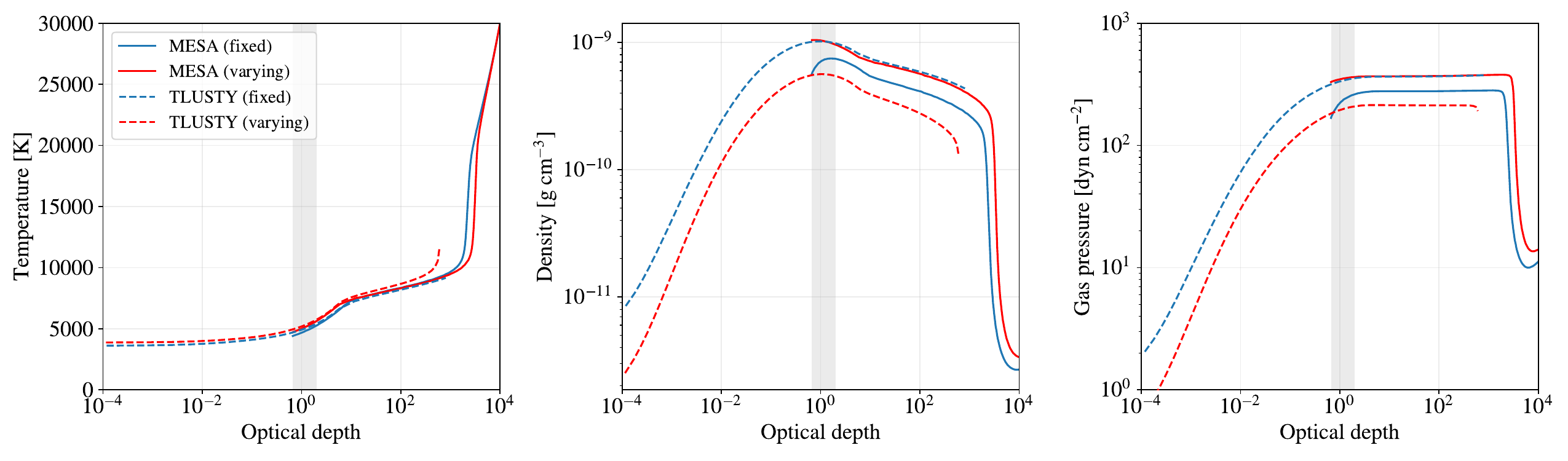}
\caption{The midlife temperature, density and gas pressure profiles of two \mesa \ models, along with \tlusty \ profiles based on their surface $(T_{\rm eff}, \log g)$. The approximate location of the spectrum-forming layers $(\tau \sim 2/3-2)$ is shaded in gray. Both \mesa \ models have $M_\star=10^6 M_\odot$ and $\alpha_{\rm MLT}=1$, differing in opacity treatment (blue for ``fixed'' and red for ``varying''). Below the photosphere, the ``varying'' \mesa \ model most closely matches the \tlusty \ model derived from the ``fixed'' \mesa \ model.}
\label{fig:TLUSTYprofiles}
\end{figure*}

We generated NLTE \tlusty \ models based on the surface conditions $(T_{\rm eff},\log g)$ of our 16 non-SR \mesa \ models, described in Table \ref{tab:grid}. As observed earlier, the surface conditions of our \mesa \ models are relatively constant throughout most of the quasi-star's lifetime, so we take the midpoint $M_{\rm BH}/M_\star=0.5$ as a representative for the surface conditions of our quasi-stars. Each of our \tlusty \ models converged to a stable point satisfying hydrostatic equilibrium, flux conservation and charge conservation. We quantify the geometry of our models with the ratio $\Delta R/R_{\rm ph}$, where $R_{\rm ph}$ is the photospheric radius computed with \mesa. $\Delta R$ is the height from the first to last layer of the \tlusty\ model,
\begin{equation}
\Delta R=\int_{m_{\rm  top}}^{m_{\rm  bottom}} \rho(m)^{-1} \mathrm{d}m\,,
\end{equation}
where $\rho$ is the density and $m$ is the column mass per unit area. In general, our tested models have consistently had $\Delta R/R_{\rm ph} < 1\%$, indicating they are well-approximated by our plane-parallel assumption.

$\alpha_{\rm MLT}$ is set in \tlusty \ via the \texttt{HMIX0} parameter. For the sake of rigor, we aimed to match it with the value used for the corresponding \mesa \ model. However, we found that models with higher values of \texttt{HMIX0} were generally easier to converge than those with lower values. As a result, a few of our \tlusty \ models do not match the mixing length parameters of their \mesa \ models (see Table \ref{tab:grid}). Fortunately, the value of \texttt{HMIX0} has little influence on our spectra, in contrast to its significance for the \mesa \ model. This is because convection becomes inefficient near the surface of a quasi-star, especially in the spectrum-forming layers; indeed, what matters for the spectrum is the $(T_{\rm eff},\log g)$ pair, and this is provided by \mesa \ models that have the intended $\alpha_{\rm MLT}$.

In Figure \ref{fig:TLUSTYprofiles}, we compare the midlife structures of two \mesa \ models with \tlusty \ models derived from their surface conditions. Both \mesa \ models were run with $M_\star=10^6 M_\odot$ and mixing length $\alpha_{\rm MLT}=1$, differing only in opacity treatment (the ``fixed'' one being the fiducial model). In all cases, the temperature profiles between the \mesa \ and \tlusty \ models agree closely. However, the density and gas pressure of the \mesa \ profiles generally differ from their \tlusty \ counterparts, in opposite directions based on the opacity treatment: ``varying'' \mesa \ models are denser than \tlusty \ by a factor of $\sim 1.5-2$, whereas ``fixed'' \mesa \ models are about half as dense as \tlusty. Interestingly, the ``fixed'' \tlusty \ models tend to closely agree with the ``varying'' \mesa \ models for our $\alpha_{\rm MLT}=1$ models, although this tendency does not hold as well for $\alpha_{\rm MLT}=2$. Whether this is a coincidence or not, it is convenient for the purposes of analysis: rather than only having \tlusty \ models that match the surface conditions of our \mesa \ models, we have \tlusty \ models that agree with the ``varying'' \mesa \ models at depths where the diffusion approximation becomes accurate. The ``fixed'' \tlusty \ models can therefore be intuitively viewed as realistic extensions to the ``varying'' \mesa \ models.

\begin{figure*}[ht!]
\centering
\includegraphics[width=\linewidth]{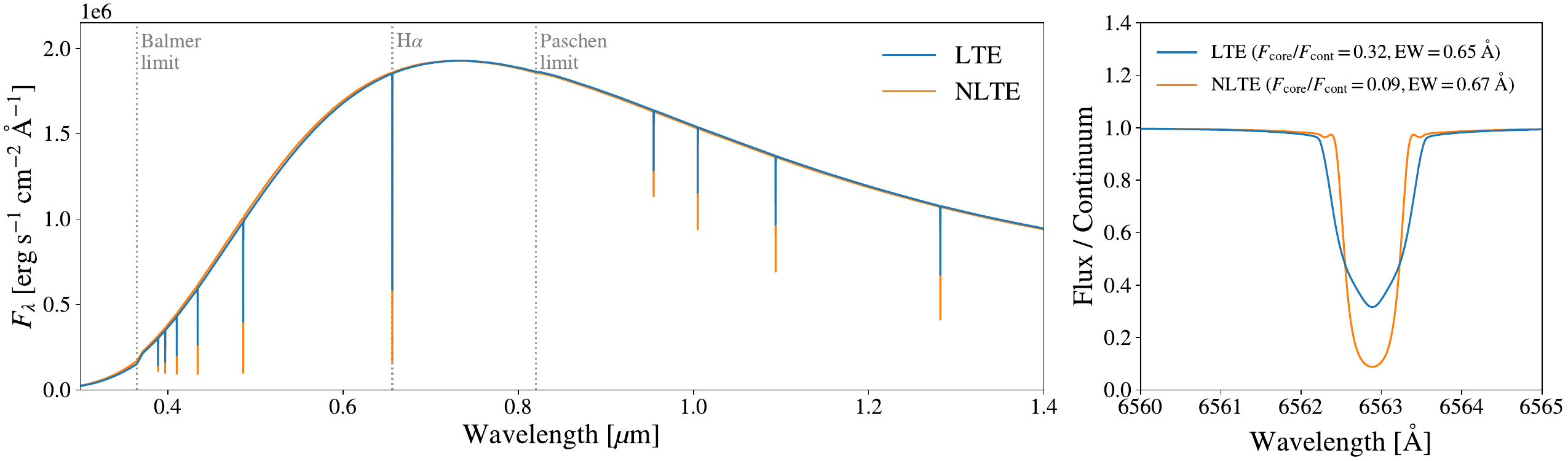}
\caption{Left: Spectrum of the fiducial NLTE model, along with an LTE model at its same conditions. Right: zoomed in comparison of their $H\alpha$ lines.}
\label{fig:LTE_vs_NLTE}
\end{figure*}

In Figure \ref{fig:LTE_vs_NLTE}, we compare the spectra of the fiducial NLTE model with its LTE counterpart, both across the Balmer and Paschen ranges as well as zoomed onto the H$\alpha$ line. In general, all of our models exhibit absorption in Balmer and Paschen lines, along with a clear Balmer break. Following \citet{deGraaf2025}, we measure the strength of the Balmer break by computing the ratio of the flux averages over [3620, 3720] Å and [4000, 4100] Å, respectively. On average, the flux ratio was $2.53\pm 0.16$ for our grid of 16 non-SR spectra; in general, ``fixed'' models had higher flux ratios than their ``varying'' counterparts, and lowering $\alpha_{\rm MLT}$ increased this ratio. While the incorporation of NLTE effects has a minor influence on the continuum shape, it substantially influences the absorption lines, which generally reach $\sim 35-55\%$ deeper than their LTE counterparts. These NLTE lines are surrounded by small emission shoulders, which can extend slightly above the continua for some of our models.

\begin{figure*}[ht!]
\centering
\includegraphics[width=\linewidth]{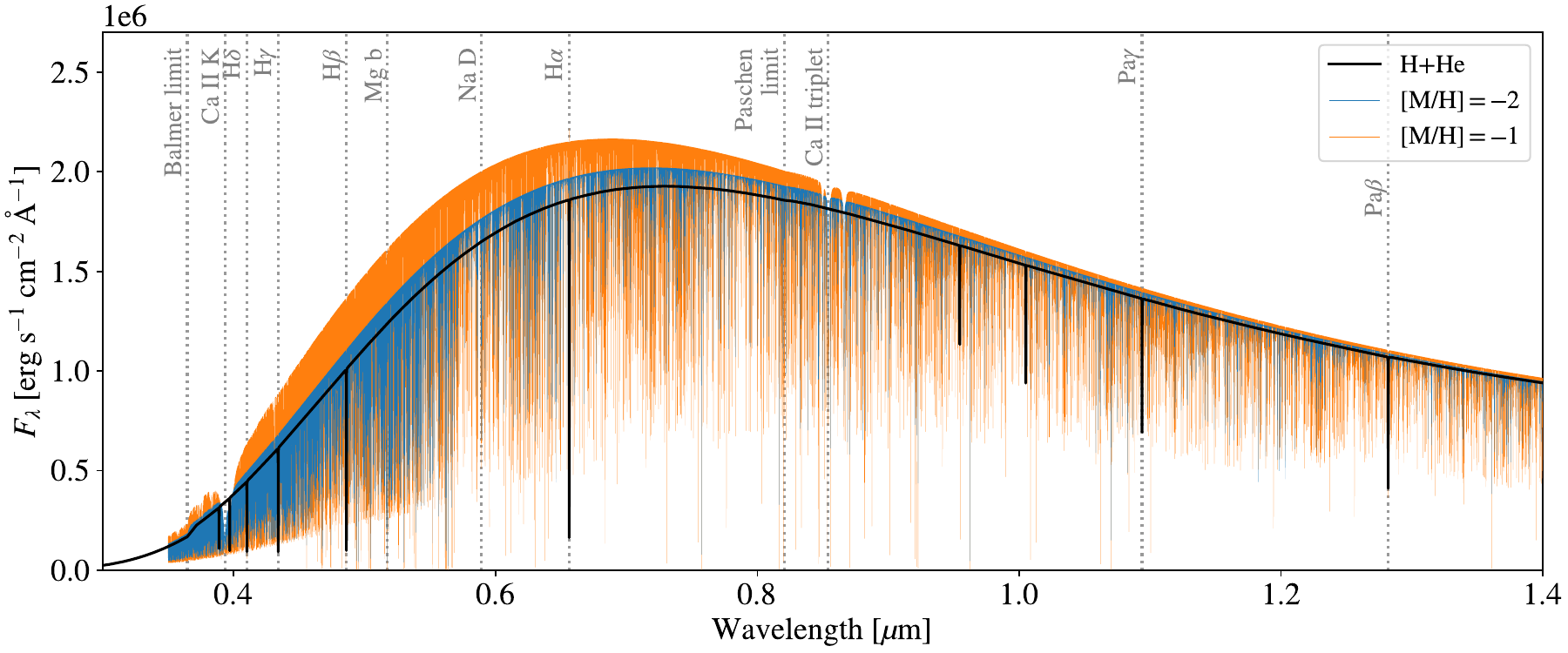}
\caption{The spectra for the same $10^6 M_\odot$ model with various metallicities.}
\label{fig:metal_compare}
\end{figure*}

In Figure \ref{fig:metal_compare}, we compare our zero-metallicity fiducial model with analogous models at $\mathrm{[M/H]}=-2$ and $\mathrm{[M/H]}=-1$. As the surface conditions of the \mesa \ models remained virtually unchanged across these metallicities, we produced the spectra for this comparison at the same $(T_{\rm eff}, \log g)$. For the sake of numerical stability, the $\mathrm{[M/H]}=-2$ and $-1$ models were computed using \texttt{HMIX0} = 8 and 7, respectively. We note that the models with nonzero metallicity were obtained using a partial opacity table, which is currently only supported in the Rybicki scheme, although our zero-metallicity spectra were nearly identical between Rybicki and CL. We find that higher metallicities correspond to brighter spectra.

Three-dimensional models are expected to suppress the surface density inversion through mixing \citep{Begelman2025}, so we also tested how SR's change to the surface conditions affects the spectrum. Since \tlusty\ takes only $(T_{\rm eff}, \log g)$ and composition from \mesa, this test probes SR through those two quantities, not through the inversion itself. We produced a NLTE \tlusty\ model at the surface conditions of our SR \mesa\ model ($T_{\rm eff} = 5085$~K, $\log g = -1.94$). As discussed earlier, this model does not feature the sudden temperature drop and density increase near the surface of our non-SR model. Due to its much lower surface gravity, modeling with \tlusty \ proved more challenging; while we were able to produce a stable atmosphere that satisfied charge and energy conservation to within 1\%, we needed to set \texttt{HMIX0} = 32 for numerical stability. This is not consistent with the $\alpha_{\rm MLT}=1$ used for the SR \mesa \ model, but as noted earlier, the mixing length appears to have very little influence on the structure at these optical depths, especially in the line-forming region. We therefore expect the spectrum to be quite similar to what $\alpha_{\rm MLT} = 1$ would yield.

\begin{figure*}[ht!]
\centering
\includegraphics[width=\linewidth]{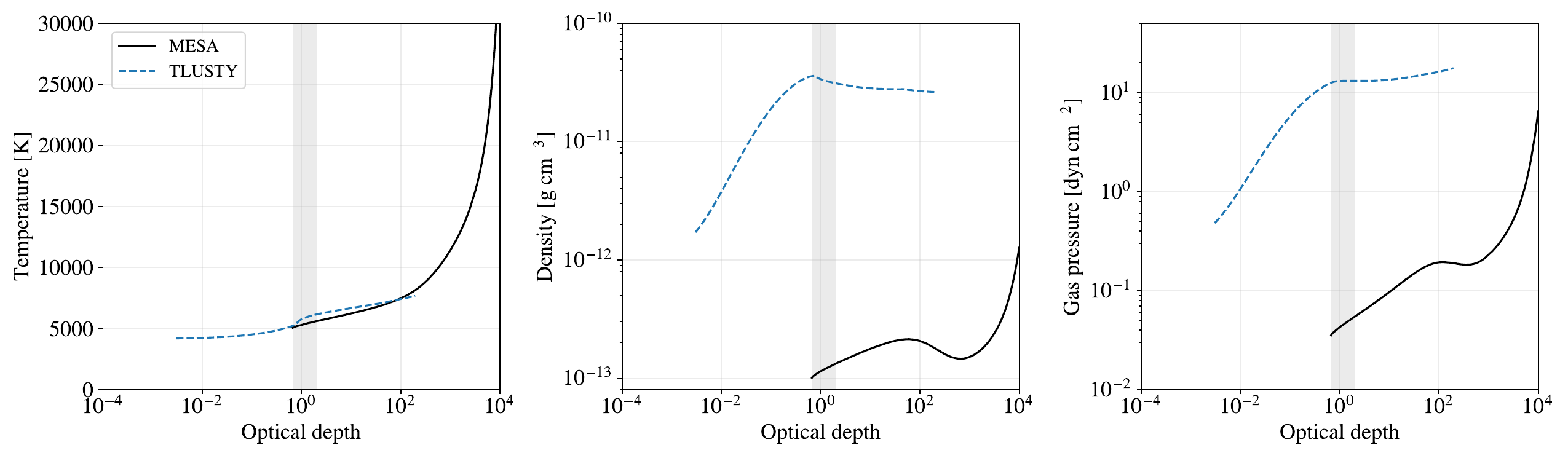}
\caption{The mid-life profile of our SR \mesa\ model, along with the \tlusty\ profile based on its surface conditions. The \mesa\ model uses $\alpha_{\rm MLT}=1$, whereas the \tlusty\ model uses \texttt{HMIX0} = 32 for numerical stability.}
\label{fig:TLUSTYprofilesSR}
\end{figure*}

\begin{figure*}[ht!]
\centering
\includegraphics[width=\linewidth]{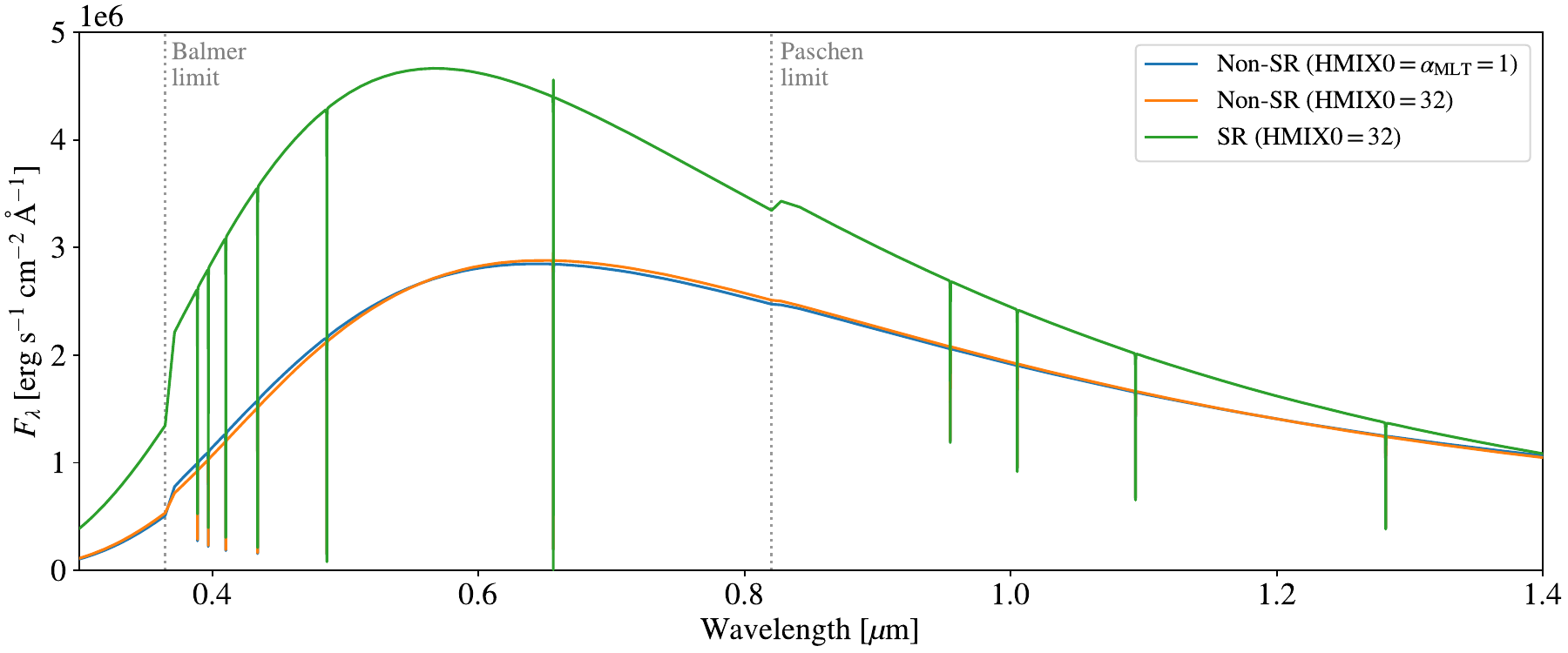}
\caption{Spectra for the NLTE \tlusty \ models based on our SR and non-SR $10^6 M_\odot$ \mesa \ models. We additionally include a non-SR \texttt{HMIX0} = 32 model to illustrate the insignificance of the \texttt{HMIX0} value on our spectra.}
\label{fig:SRspectrum}
\end{figure*}

Figure \ref{fig:TLUSTYprofilesSR} compares the atmospheric structure of the SR \mesa \ model with the \tlusty \ model based on it. As with the non-SR case, our \tlusty \ temperature profile agrees closely with that of the \mesa \ model, in spite of the differences in $\alpha_{\rm MLT}$. However, the deviations in the density and gas pressure are far greater, with the \tlusty \ models being over 100 times denser than the corresponding \mesa \ models. Such a deviation is unlikely to be due to the \texttt{HMIX0} discrepancy: our non-SR fiducial model with \texttt{HMIX0} = 32 is only about twice as dense as its \mesa\ counterpart. The difference is likely not numerical. At these surface conditions, the photospheric opacity is dominated by electron scattering off the small ionized fraction of hydrogen, which decreases with density. Because $\sigma T_{\rm eff}^4/(cg) \approx 110$, the Eddington factor $\Gamma \approx 110\,\kappa$ is set by the density, and hydrostatic equilibrium admits two photospheric solutions at the same $(T_{\rm eff}, \log g)$: a low-density, radiation-supported one with $\Gamma \approx 1$, and a denser, gas-supported
one with $\Gamma \ll 1$. The SR \mesa\ model lies on the first ($\Gamma_{\rm surf} \approx 0.96$, $P_{\rm gas}/P \approx 0.02$), set by the radiation-dominated envelope beneath it; \tlusty, which has no envelope beneath it, converges to the second. For the non-SR models ($\Gamma_{\rm surf} \sim 0.01$), only the gas-supported solution exists, and the two codes agree to within a factor of two. 

In Figure \ref{fig:SRspectrum}, we compare the spectra with and without SR. For the non-SR model, we include spectra computed with \texttt{HMIX0} = 1 and 32 in \tlusty, which are nearly identical. The SR spectrum shows the same features as the non-SR one: a blackbody-like continuum and strong absorption lines. However, it is the spectrum of a gas-supported atmosphere at the SR surface conditions, not of the SR \mesa\ photosphere. A photosphere at $\Gamma \approx 1$ is unlikely to be static, and modeling it consistently would require an
outflowing atmosphere.

\subsection{Comparison with LRDs}
\begin{figure*}[ht!]
\centering
\includegraphics[width=\linewidth]{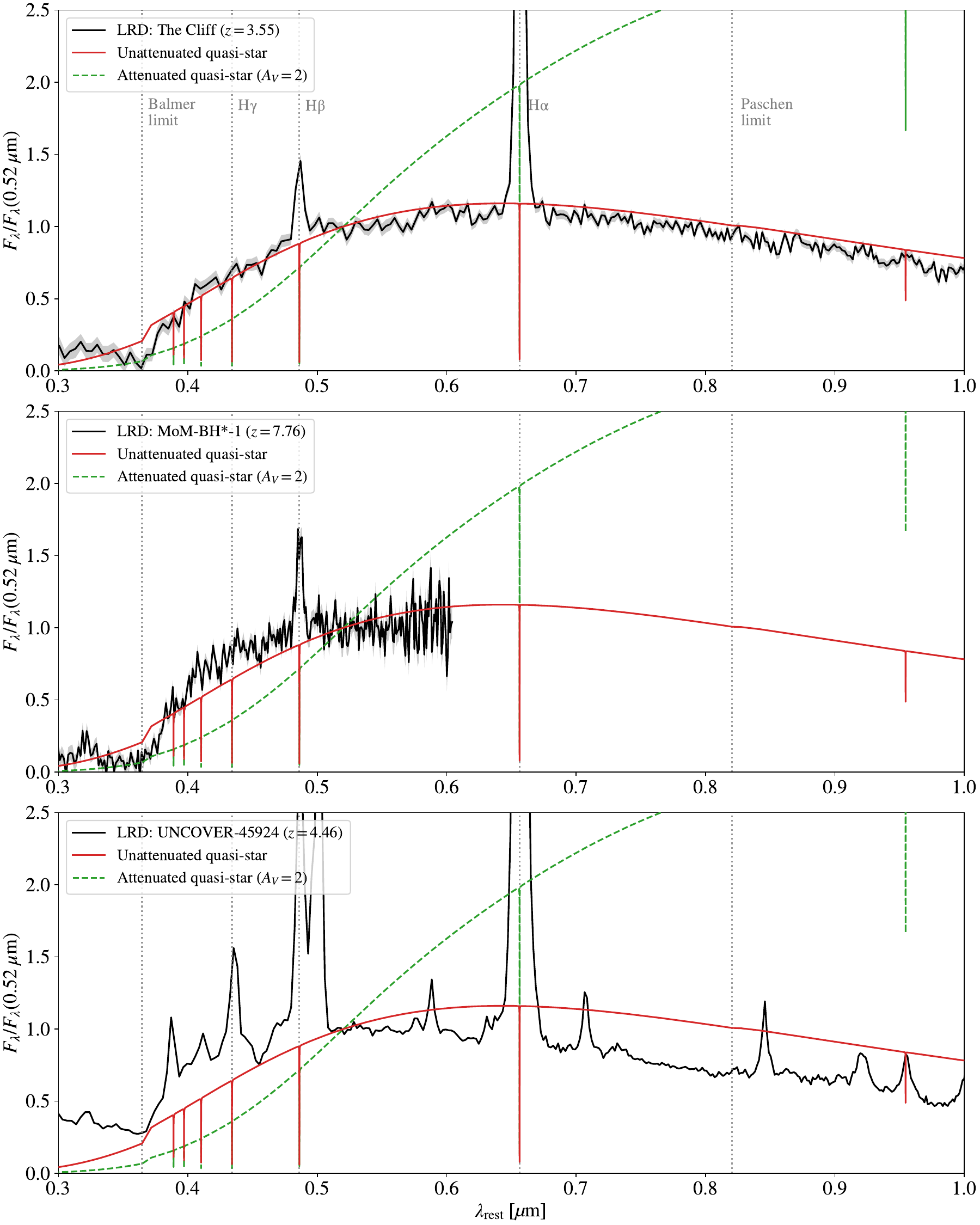}
\caption{Comparison of several LRDs with the midlife spectrum of a ``varying'' $10^6 M_\odot$ model with $\alpha_{\rm MLT}=1$, both with and without attenuation ($A_V = 2$). Flux is normalized at $0.52 \mu \mathrm{m}$. Strong attenuation generally results in a poorer match between our quasi-star spectrum continua and the observations from LRDs.}
\label{fig:attenuation}
\end{figure*}

In Figure \ref{fig:attenuation}, we compare the midlife spectrum of a ``varying'' $10^6 M_\odot$ model with the three LRDs analyzed in \cite{Santarelli2025}: The Cliff \citep{deGraaf2025}, MoM-BH*-1 \citep{Naidu2026}, and UNCOVER-45924 \citep{Labbe2024}. We normalize the flux at $0.52 \mathrm{\mu m}$ to study the spectral shape independent of the particular luminosity. As a point of comparison, we also include an attenuated version utilizing the dust curve from \citet{Gordon2003}, as was done by \cite{Martins2026}. We use a normalization of $A_V=2$ for this curve. We do not subtract the contributions from the host galaxy from these LRDs, due to the lack of clear data on those contributions.

Without attenuation, the continua of our models broadly agree with those of the LRDs. The average Balmer break strength of $2.53\pm 0.16$ across our non-SR models is typical of LRDs \citep{Sun2026}, albeit lower than extreme cases like The Cliff's $6.9^{+2.8}_{-1.5}$ \citep{deGraaf2025}. Attenuation generally worsens the spectral resemblance, raising the redder side of the spectrum and diminishing the Balmer break. A major difference is that our quasi-star spectra feature Balmer and Paschen absorption lines, an aspect we found to be strengthened due to the incorporation of NLTE effects. While LRDs can contain absorption dips in their lines, they generally feature pronounced emission lines with high Balmer decrements.

The absence of emission lines does not preclude quasi-stars from being LRDs, as there are multiple potential sources for them not accounted for by \tlusty. \cite{Martins2026} found that emission lines can be produced by mass loss, something likely to exist for quasi-stars; however, this came at the expense of the strength of the Balmer break, typically leading to the formation of a Balmer jump. 
Additionally, while we produced spectra corresponding to SR, \mesa \ and \tlusty \ are nonetheless both 1D codes, and there are limitations in their abilities to model convection and porosity.
More generally, \citet{Chang2026multiphase} demonstrated that the relative strengths of the Balmer lines and Balmer break cannot be accounted for by a single population of hydrogen atoms, and thus a multiphase, clumpy gas structure is necessary in any viable LRD models. Our extensive investigation has led us to the same conclusion.

\begin{figure*}[ht!]
\centering
\includegraphics[width=\linewidth]{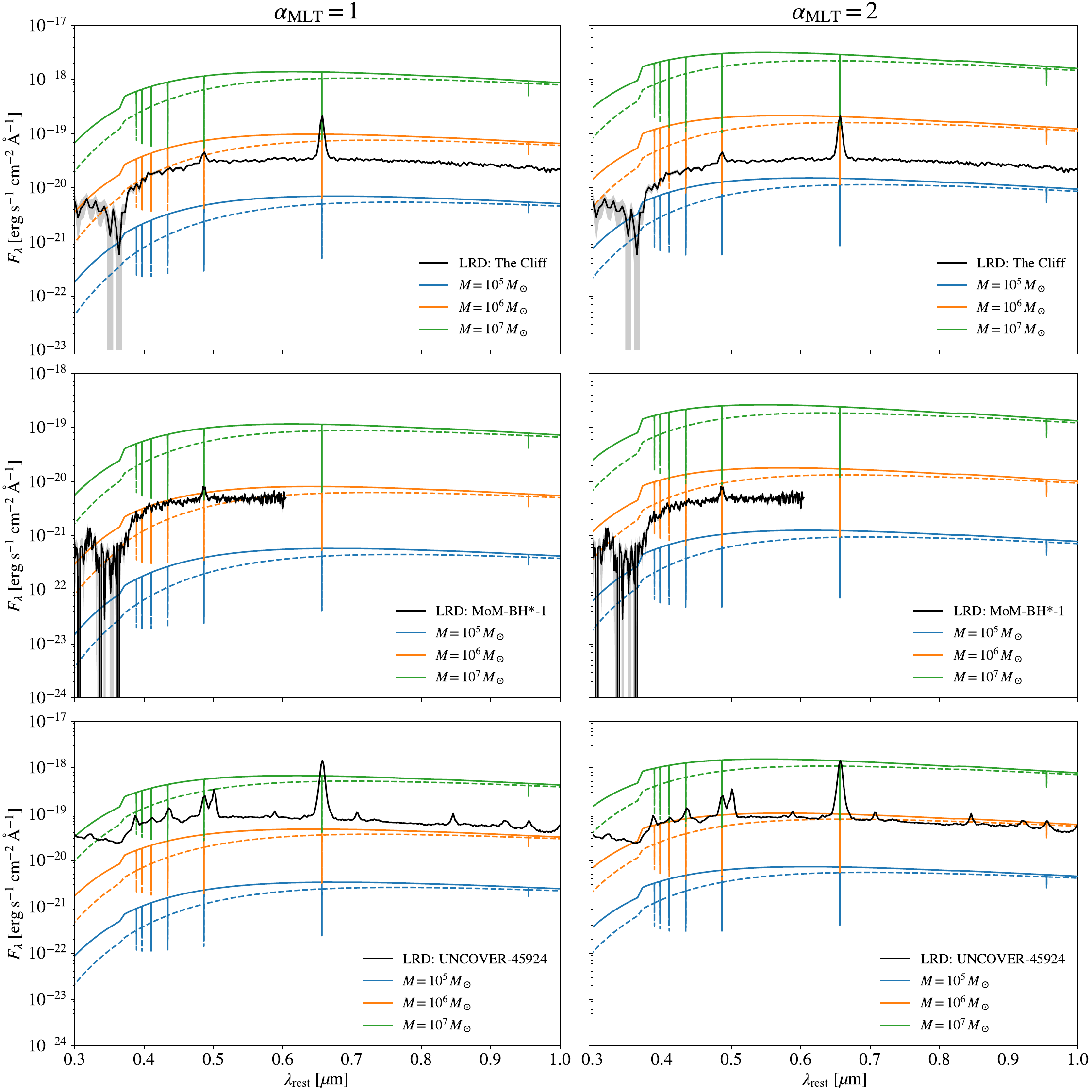}
\caption{Comparison of LRDs with midlife quasi-star models, without flux normalization. Results are shown for both $\alpha_{\rm MLT} = 1$ and 2, as well as with ``varying" (solid) and ``fixed" (dashed) opacity treatments.}
\label{fig:spectra_masses}
\end{figure*}

Our spectra also provide us a means to estimate the masses of LRDs, if they are indeed quasi-stars. In Figure \ref{fig:spectra_masses}, we compare the aforementioned LRDs with multiple non-SR quasi-star models of different masses, without flux normalization. We note that the effective temperature, and therefore the flux, is influenced by the mixing length parameter $\alpha_{\rm MLT}$ and the treatment of opacity. Despite this uncertainty, we can still infer that The Cliff and MoM-BH*-1 are generally bounded between our $10^5 M_\odot$ and $10^6 M_\odot$ quasi-star models, whereas UNCOVER-45924 lies between $10^6-10^7 M_\odot$. More generally, the mass of an observed quasi-star can be inferred from its spectrum if the treatment of $\alpha_{\rm MLT}$ and opacity are established. It should be noted that the spectra in these figure are taken at the time when $M_{\rm BH}/M_\star = 0.5$, although the effective temperature remains relatively constant between $0.25 \leq M_{\rm BH}/M_\star \leq 0.75$. As such, the order of magnitude for a quasi-star's mass can still likely be inferred regardless of our uncertainty in the quasi-star's age.

\section{Summary} \label{sec:summary}
We have presented significant developments in the evolutionary modeling of quasi-stars with \mesa, the first non-LTE atmospheric models based on numerical simulations of quasi-stars from first principles, and the first successfully converged non-LTE \tlusty \ models at the conditions and compositions expected of quasi-stars. We have provided modeled spectra for a multitude of quasi-star evolutionary constructions, including different compositions, mixing length parameters, opacity treatments, and a model with the surface density inversion numerically suppressed.

Our main results can be summarized as follows:
\begin{itemize}
    \item The upper limit on the BH mass for older models (CB24, \citealt{Hassan2026}) appears to be caused by the enforcement of saturated convection out to high radii. Our current model, which utilizes saturated convection for the luminosity near the BH but models the rest of the quasi-star with \mesa's \ mixing length theory, can describe the full lifetime of the quasi-star up to $M_{\rm BH}/M_\star \sim 1$.
    \item The surface conditions of our \mesa \ model, and thus the spectra based on them, change little across most of the quasi-star's evolution. They are primarily determined by the mass of the quasi-star, along with physical decisions such as the choice of $\alpha_{\rm MLT}$ and treatment of opacity.
    \item The incorporation of NLTE effects into our quasi-star spectra strengthen Balmer line absorption by a factor on the order of $\sim 35-55\%$.
    \item The features of our quasi-star spectra generally agree with the continua observed in LRDs. Our models have blackbody-like continua and Balmer breaks typical of LRDs in the literature. The remaining discrepancy is that our models feature Balmer and Paschen absorption rather than the observed emission lines.
    \item The order of magnitude for the masses of the LRDs we analyzed, if they are indeed quasi-stars with $1\leq\alpha_{\rm MLT}\leq2$, can be inferred to lie within $\log_{10} (M_\star/M_\odot)=6\pm1$.
\end{itemize}

Our results represent the most rigorous comparison of quasi-stars with LRDs to date. However, further investigations will be needed to overcome the current limitations of our modeling. This includes finding a method of making \mesa \ completely consistent with the saturated flux limit at all depths, modeling layers above the photosphere that may contribute to Balmer emission, and a detailed study of how non-isotropic accretion and winds at a quasi-star's surface can influence its spectrum.

\begin{acknowledgments}
JBH and RP acknowledge support from NASA award 80NSSC25K7554.
JBH and PJA acknowledge support from award 644616 from the Simons Foundation.  MCB acknowledges support from NASA Astrophysics Theory Program grant 80NSSC24K0940.
The Center for Computational Astrophysics at the Flatiron Institute is supported by the Simons Foundation.

Claude Code was utilized to identify a sequence of steps for obtaining our converged \tlusty \ models, as well as to improve the code and find stabilizing settings for the \mesa \ models.
\end{acknowledgments}

\bibliography{biblio}

\end{document}